\documentclass[12pt,preprint]{aastex}
\newdimen\minuswidth    %define @ width of minus sign for tables
\setbox0=\hbox{$-$}
\minuswidth=\wd0
\catcode`@=\active
\def@{\kern\minuswidth}
\newdimen\digitwidth    %define ! a one digit width for tables
\setbox0=\hbox{\rm0}
\digitwidth=\wd0
\catcode`!=\active
\def!{\kern\digitwidth}

\def\alwaysmath#1{\ifmmode{#1}\else{$#1$}\fi}
\def\kms{\alwaysmath{{{\rm km\,s^{-1}}}}}
\begin{document}
%\slugcomment{Submitted to ApJ}

\shorttitle{C- and O-depleted BSS in 47 Tuc}
\shortauthors{Ferraro et al.}

\title{Discovery of Carbon/Oxygen depleted Blue Straggler Stars in 47
  Tucanae: the chemical signature of a mass-transfer formation
  process}
\footnotetext[1]{Based on observations 
  collected at the ESO-VLT (Cerro Paranal, Chile) under programme 072.D-0337.}
\author{
F.R. Ferraro\altaffilmark{2},   
E. Sabbi\altaffilmark{3},  
R. Gratton\altaffilmark{4},
G. Piotto\altaffilmark{5}, 
B. Lanzoni\altaffilmark{6}, 
E. Carretta\altaffilmark{6}, 
R.T. Rood\altaffilmark{7}, 
Alison Sills\altaffilmark{8}, 
F. Fusi Pecci\altaffilmark{6}, 
S. Moehler\altaffilmark{9,10}, 
G. Beccari\altaffilmark{6,11,12}, S. Lucatello\altaffilmark{4},
N. Compagni\altaffilmark{2}}
\footnotetext[2]{Dipartimento di Astronomia Universit\`a di Bologna, 
via Ranzani 1, I--40127 Bologna, Italy,  francesco.ferraro3@unibo.it}
\footnotetext[3]{Space Telescope Science Institute, 3700 San 
Martin Drive, Baltimore, MD, 21218, USA}
\footnotetext[4]{INAF--Osservatorio Astronomico di Padova, 
Vicolo dell'Osservatorio 5, I--35122, Padova, Italy}
\footnotetext[5]{Dipartimento di Astronomia, Universit\`a di Padova, 
Vicolo dell'Osservatorio 2, I-35122 Padova, Italy}
\footnotetext[6]{INAF-Osservatorio Astronomico di Bologna, 
via Ranzani 1, I-40127 Bologna, Italy}
\footnotetext[7]{Astronomy Department, University of Virginia, 
P.O. Box 3818, Charlottesville, VA, 22903--0818}
\footnotetext[8]{Department of Physics and 
Astronomy, McMaster University, 1280 Main Street West, Hamilton, ON L8S 4M1, Canada}
\footnotetext[9]{Institute f\"{u}r Theoretische Physik und Astrophysik
 der Universit\"{a}t Kiel, D--24098 Kiel, Germany}
\footnotetext[10]{European Southern Observatory, Karl Schwarzschild Strasse 2,
D-85748, Garching bei Munchen, Germany}
\footnotetext[11]{Dipartimento di Scienze della Comunicazione,
Universit\`a degli Studi di Teramo, Italy}
\footnotetext[12]{INAF--Osservatorio Astronomico
di Collurania, Via Mentore Maggini, I--64100 Teramo, Italy}
%\date{18 June 06}

\begin{abstract}  
%We use high-resolution spectra obtained with the ESO Very Large Telescope to
%probe the formation process of Blue Stragglers stars (BSS) in Galactic
%globular clusters. Basic physical properties, such as rotational and radial
%velocities and surface abundance patterns, have been obtained for 43 BSS in
%47 Tuc. 
We use high-resolution spectra obtained with the ESO Very Large Telescope
to measure surface abundance patterns of 43 Blue Stragglers stars (BSS)
in 47 Tuc.
We discovered that a sub-population of BSS shows a significant
depletion of Carbon and Oxygen with respect to the dominant population.  This
evidence would suggest the presence of CNO burning products on the BSS
surface coming from a deeply peeled parent star, as expected in the case of
mass-transfer process. This is the first detection of a chemical signature
clearly pointing to a specific BSS formation process in a globular cluster.
\end{abstract}  


\section{Introduction}   
\label{intro}  
Globular Clusters (GCs) are important astrophysical laboratories for
studying the evolution of single stars as well as binary systems. In
particular, the evolution and the dynamical interactions of binary
systems in high-density environment can generate objects that cannot
be explained by standard stellar evolution (like X-ray binaries,
millisecond pulsars, etc.).

In this respect the most common by-product of binary evolution are the
so-called Blue Straggler Stars (BSS). They are commonly defined as those
stars brighter and bluer (hotter) than the main sequence (MS) turnoff (TO)
stars. BSS lie along an extrapolation of the MS, and thus mimic a rejuvenated
stellar population. First discovered by \citet{sandage53} in M3, their nature
has been a puzzle for many years, and their formation mechanism is still not
completely understood. BSS are more massive than the normal MS stars
\citep{shara97}, thus indicating that some process which increases the
initial mass of single stars must be at work.  Such effects could be related
either to mass transfer between binary companions, the coalescence of a
binary system or the merger of two single or binary stars driven by stellar
collisions.

According to \citet[][see also Davies, Piotto \& de Angeli 2004]{fusi92}, BSS
in different environments could have different origins. In particular, BSS in
loose GCs might be produced from coalescence/mass-transfer of primordial
binaries (hereafter BMT-BSS), whereas in high density GCs 
%(depending on survival-destruction
%rates for primordial binaries) 
BSS might arise mostly from stellar
collisions (C-BSS), particularly those that involve binaries. As shown by
\citet{ferraro03}, both formation channels can be very efficient in producing
BSS in their respective typical environment (see the case of M80 and NGC288).
%Ferraro et al. 1999, Bellazzini et al. 2002). 
Moreover, there is evidence
that different formation mechanisms could also act simultaneously within the
same cluster in different radial regions corresponding to widely different
stellar densities. This is suggested by the bimodal BSS radial distribution
observed in a few clusters (M3, 47 Tuc, NGC6752, M55 and M5) where the BSS
specific frequency ($F_{BSS}$) has been found to be highly peaked in the cluster center,
rapidly decreasing at intermediate radii and rising again outward (see
Ferraro 2006 for a review).

While theoretical models still predict conflicting results on the expected
properties of BSS generated by different production channels, systematic
spectroscopic observations have recently begun to provide the first set of
basic properties (mass, rotation velocities, etc.; see the recent work by De
Marco et al. 2005). However, with the exception of a few bright BSS in the
open cluster M67 \citep{mathy91, she00}, an extensive survey of BSS surface
abundance patterns is still lacking, particularly in GCs.

The advent of 8-meter class telescopes equipped with multiplexing capability
spectrographs allows us to perform extensive surveys of surface abundance
patterns for representative numbers of BSS in GCs, thus filling for the very
first time such a gap in our knowledge. We are currently performing such a
survey for a sample of galactic GCs with different characteristics, by using
the Very Large Telescope (VLT) of the European Southern Observatory (ESO). In
this letter we present the first results, concerning the discovery of a
population of C- and O-depleted (hereafter CO-depleted) BSS in 47 Tuc.

\section{Observations}
\label{obs}

A sample of 43 BSS in the magnitude range $V\sim 15.5 -16.8$ was observed
over almost the entire radial extent of 47 Tuc (from $\sim 20''$ up to $\sim
16'$ from the cluster center).  The observations were performed at the
ESO-VLT, during four nights in October 2003, by using the multiplex facility
FLAMES-GIRAFFE in MEDUSA mode, which allows the simultaneous observation of
up to 130 targets within a $\sim 25'$-diameter field of view.  With the main
aim of determining C, O, two different set-ups were used: $HR22$ (R=11700,
$\lambda_c= 9205\,$\AA ), $HR18$ (R=18300, $\lambda_c= 7691$\AA), thus
sampling the C{\,\sc i} line at $\lambda = 9111.8$\AA and the O{\,\sc i}
triplet at $\lambda \sim 7774\,$\AA, respectively.
% FLAMES-GIRAFFE in MEDUSA mode. This combination mode allows the simultaneous
% observation of (up to) 130 targets within a $\sim 25'$-diameter field of
% view. Three different set-ups were used: (1) $GIRAFFE,\ HR22$ (R=11700),
% centered at $\lambda 9205$ \AA\ ; (2) $GIRAFFE,HR18$ (R=18300), centered
% at $\lambda 7691$ \AA\ and (3) $GIRAFFE,\ HR19$ (R=13300), centered at
% $\lambda 8053$ \AA, in order to observe the C{\,\sc i}\ line at $\lambda =
% 9111.8$ \AA, the O {\,\sc i} triplet at $\lambda \sim 7774$ \AA, and the Na
% {\,\sc i} doublet at $\lambda = 8183-8194$ \AA, respectively.

All the targets were selected from the   photometric 
%and astrometric
catalog published by \citet{ferraro04}, choosing only relatively isolated
objects. Two different pointings were performed in order to sample the inner
and the most external regions of the cluster. Our ability to observe BSS in
the inner 100\arcsec\ was greatly restricted by the crowding of both stars
and fibers. In each pointing $\sim 30$ BSS were observed, while $\sim 20$
fibers were used to acquire sky spectra. Seventeen BSS were observed in both
the pointings and were used to test the internal accuracy of the abundance
measures.  The total exposure time of each pointing was split in
sub-exposures of about one hour each. In summary, for each of the two
pointings we have obtained 3 spectra sampling the C{\,\sc i} lines and 2 for
the O{\,\sc i}. By combining the sub-exposures we finally obtained mean
spectra with a $S/N \ga 50$ (per resolution element) for most of the selected
BSS.  Raw spectra were reduced in IRAF following the standard procedure.
%Each spectrum was
%corrected for the appropriate bias, and an average flat field was
%created and normalized. 
The task APALL was used to define and extract the apertures and to calibrate
the one-dimensional spectra in wavelength, by adopting the dispersion
solution derived by Th-Ar lamps acquired after each spectra.  A full
description of the data analysis will be given in a forthcoming paper (Sabbi
et al. 2006, in preparation).  Here we focus on the discussion of C and O
abundances.   The analysis of the chemical abundances was performed using
the ROSA package \citep{gratton88}. The equivalent width (EW) of each
measurable line was measured by Gaussian fitting of the line profile,
adopting a relationship between EW and FWHM, as described in
\citet{bragaglia01}; an iterative clipping average over a fraction of the
highest spectral points around each line was applied to define a local
continuum. Abundances were derived from the measured EW once appropriate
atmospheric parameters have been adopted. In particular: (1) stellar
temperatures ($T\simeq 6600-8000$\,K) were estimated by the empirical
relation $\log T=-0.38 { (B-V)} +3.99$, obtained from the H$_\alpha$
temperatures of TO stars and of a few BSS, observed at high resolution
($R\sim 40,000$) with UVES \citep[][Sabbi et al. 2006]{carretta05}; (2)
gravities ($\log g\simeq 4.3$--4.8) were estimated from the BSS location in
the Color Magnitude Diagram (CMD, see Fig.~\ref{f2}), and (3) a value of
2\,\kms was assumed for the microturbulence velocity.  Finally,
[Fe/H]$=-0.67$ has been adopted from \citet{carretta04}. The derived
abundances of [O/Fe] were corrected for departures from the local
thermodynamic equilibrium (NLTE), following \citet{gratton99}\footnote{ The
NLTE correction for [C/Fe] was derived by interpolating the C{\,\sc i}
abundances listed by \citet{tomkin92}. The derived expression is: ${\rm
C{\,\small I}_{NLTE}-C{\,\small I}_{LTE}}=-1.228 \times 10^{-4} T+ 5.9082
\times 10^{-2} \log g + 2.552 \times 10 ^{-3} {\rm EW}-2.125 \times
10^{-5}{\rm EW}^2+0.345 $}.
 
\section{Results}
Radial velocity (RV) measurements have shown that all the selected BSS are
likely members of 47 Tuc. In fact, the mean heliocentric velocity turns out
to be $\langle{\rm RV}\rangle=-17.1\pm1.4\,\kms$, in good agreement
with the value listed by \citet[][ 2003, $\langle{\rm RV}\rangle=-18.7\,\kms$]{harris96}.

\begin{figure}
\epsscale{1.1}   
\plotone{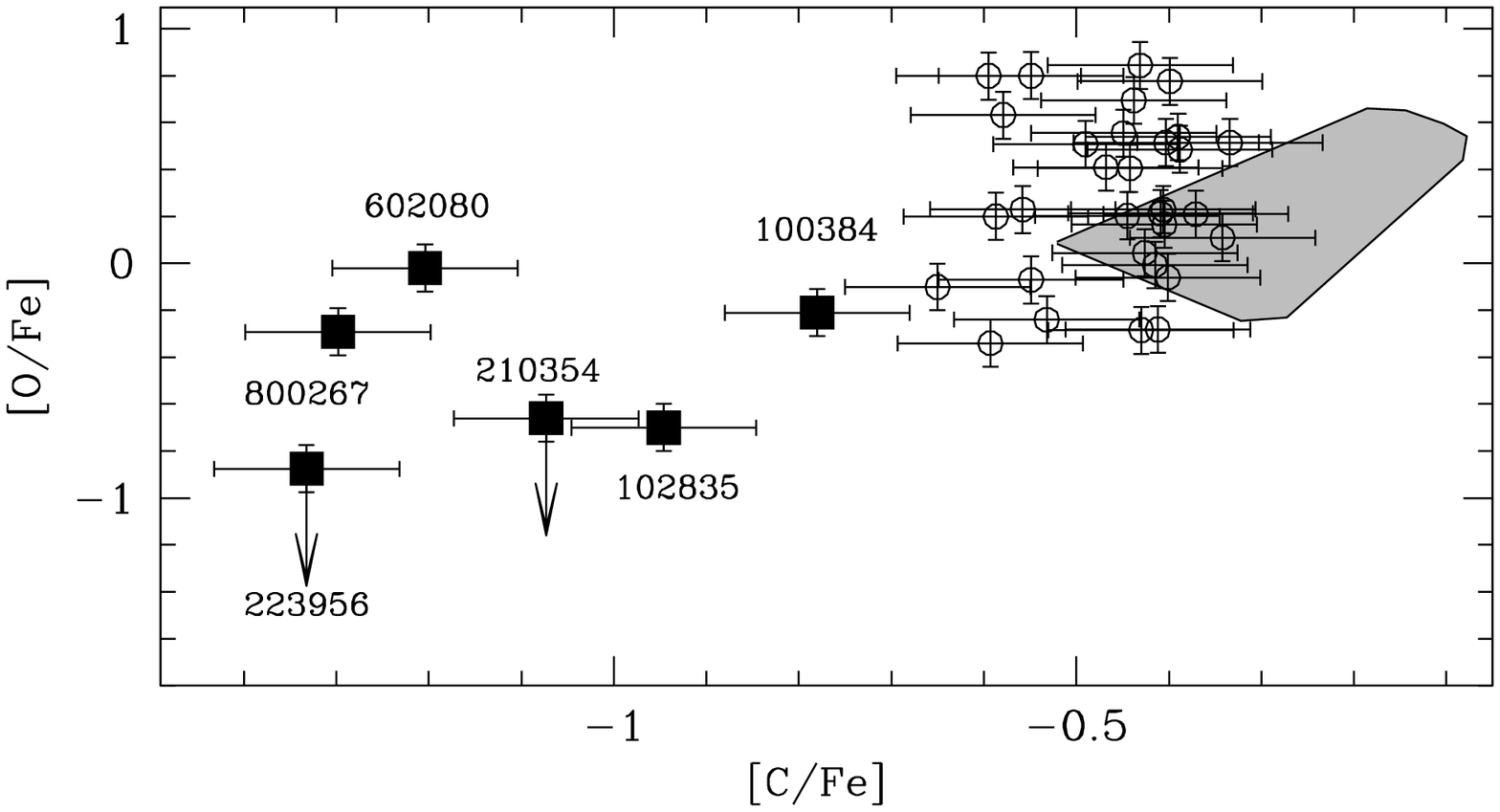}
\caption{[O/Fe] ratio as a function of [C/Fe] for the BSS observed in 47 Tuc.
Normal BSS are marked with {\it empty circles}, while CO-depleted BSS are
marked with {\it filled squares} and their names are also reported. The gray
regions correspond to the location of the 12 TO stars in 47 Tuc analyzed by
\citet{carretta05}.}
\label{f1}
\end{figure}

Figure~\ref{f1} shows the major result of the abundance analysis
%: the [Na/Fe] and [O/Fe] abundances are plotted as a function of the [C/Fe]
%abundance for all the 43 observed BSS.  As can be seen 
and reveals the presence (beside the main population) of a small group of BSS
showing abundance anomalies.
%that two main groups of BSS can be identified.  
The main group comprises the majority (36 objects) of the BSS in our sample,
showing abundances roughly in agreement with those of TO stars (gray regions)
obtained by \citet{carretta05}, with a peak\footnote{There is an offset of
about 0.2--0.3 dex in C abundances between the main group of the BSS and the
TO-stars. We have not been able to determine whether this is a real abundance
effect or due to the different indicators used for the BSS (high excitation
C{\,\sc i} lines) and for the TO-stars (CH bands).}  at ${\rm [C/Fe]}\simeq -
0.4$.  Six objects (14\% of the sample)
%populate the left side of the diagrams, with 
have ${\rm [C/Fe]}<-0.7$, i.e., C abundances 4$\sigma$ lower than the peak of
the other group distribution.  They show a large spread (1 dex) in [C/Fe] and
[O/Fe], a significant depletion of C and, though to a lesser extent, of O
abundances.  Moreover, one object in our sample (700912) shows quite large
rotation velocity ($V_{\rm rot} \sin i \sim 100 \,\kms$): no abundance
measurements were possible for this rapid rotator.
% because of the weakness of the lines and the presence of sky
%emission lines. 
The CO-depleted BSS seem to share the same radial distribution of ``normal''
BSS, and a Kolmogorov-Smirnov (K-S) test yields a probability of $53\%$ that
the two radial distributions are extracted from the same parent population.

\begin{figure}
\epsscale{1.1}   
\plotone{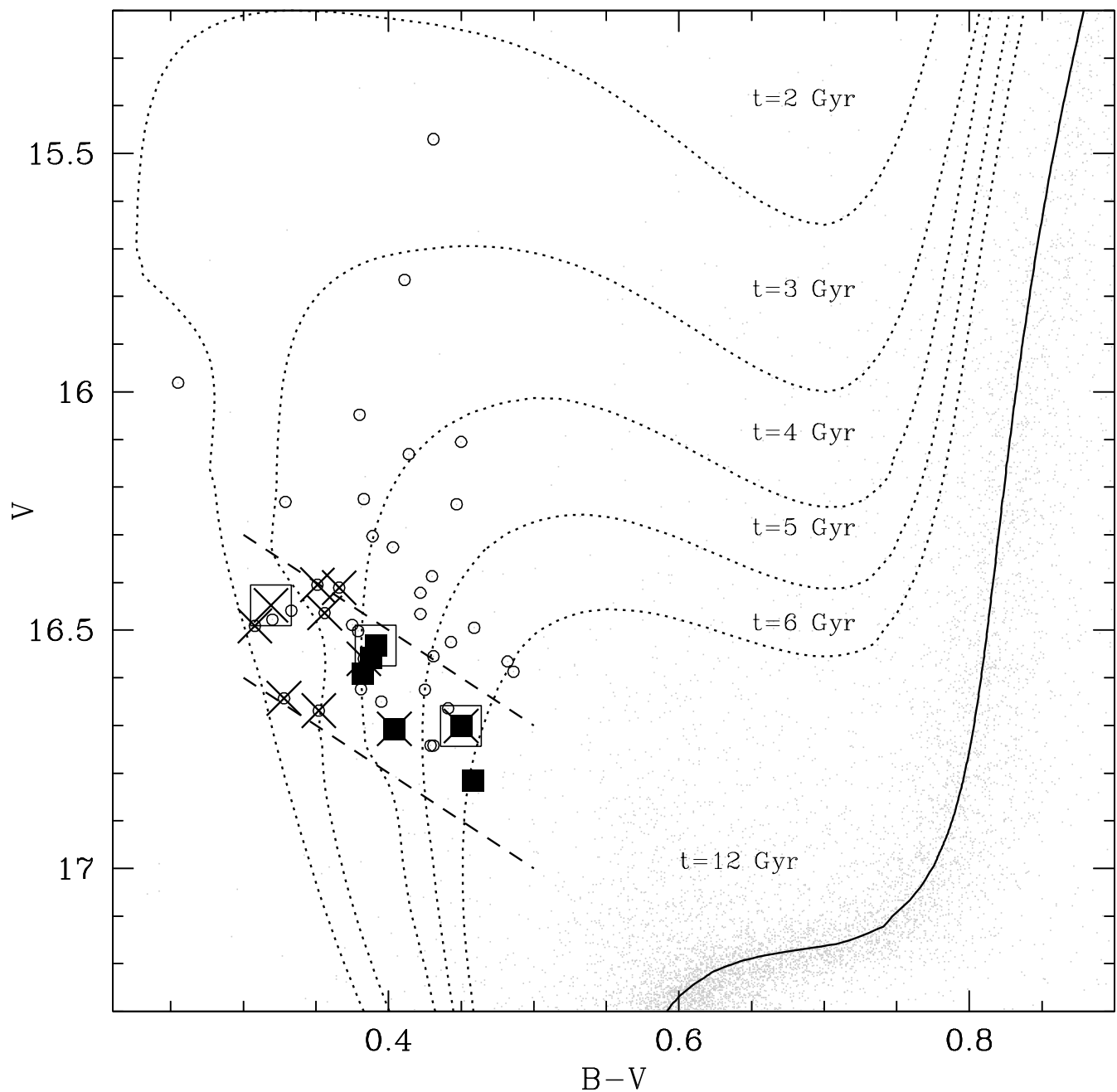}
\caption{Zoomed CMD of 47 Tuc in the BSS region. Normal BSS are marked with
{\it open circles}, while CO-depleted BSS are shown as {\it filled squares}.
Isochrones of different ages (from 2 to 12 Gyr) from Cariulo et al. (2003)
are overplotted for comparison. The three W~UMa systems and the 10 BSS
rotating with $V_{\rm rot} \sin i > 10\, \kms$ are highlighted with {\it
large empty squares} and {\it large crosses}, respectively.  }
\label{f2}
\end{figure}

\begin{figure}
\epsscale{1.0}   
\plotone{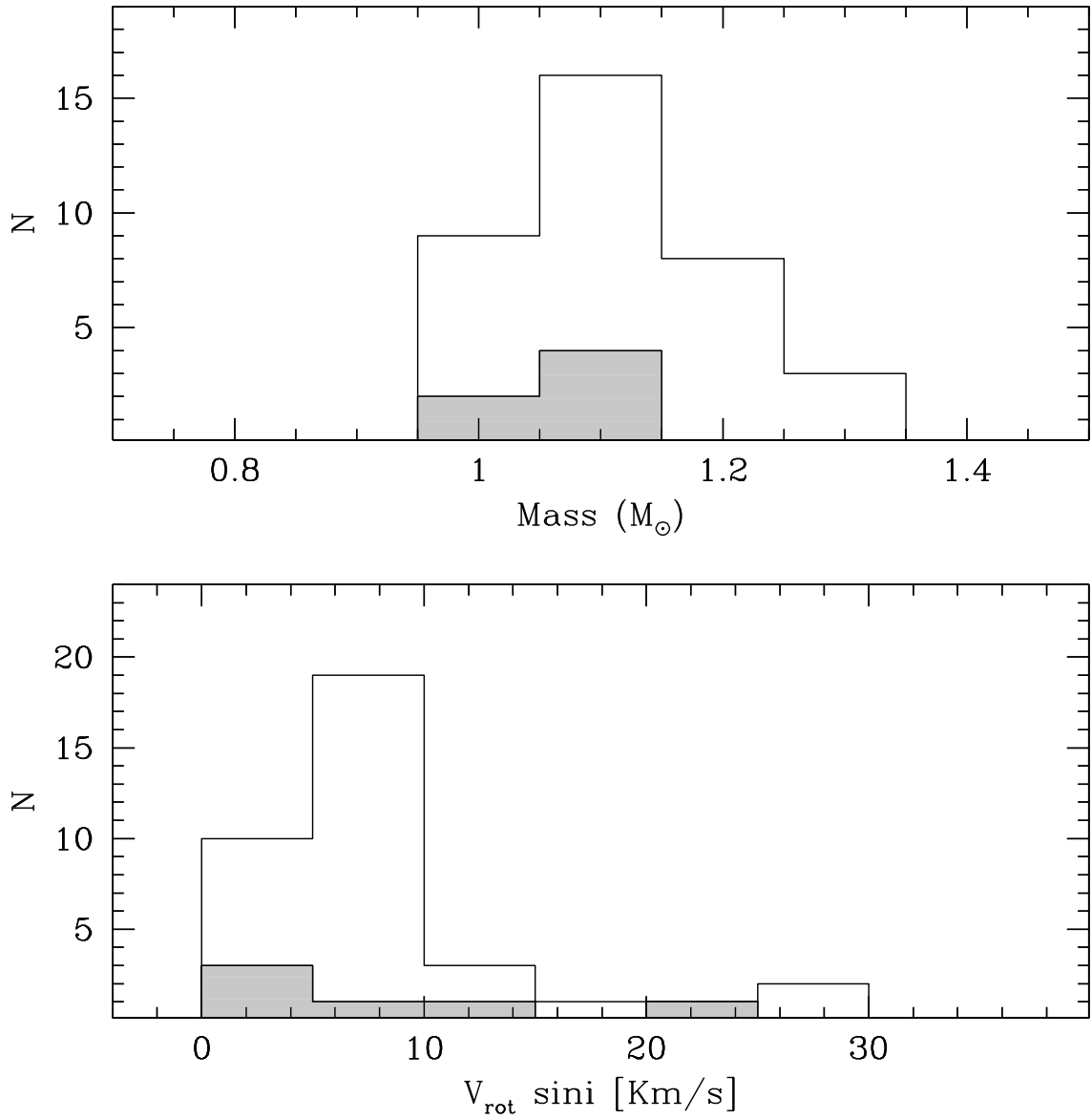}
\caption{{\it Upper panel:} The derived mass distribution of normal 
BSS ({\it empty histogram}) is compared to the CO-depleted BSS
distribution ({\it grey histogram}). {\it Lower panel:} The rotation
velocity distribution of normal BSS ({\it empty histogram}) is
compared to the CO-depleted BSS distribution ({\it grey histogram}). }
\label{f3}
\end{figure}

Figure~\ref{f2} shows the area of the CMD pertinent to the BSS region.
%, where all the selected targets are marked with the same symbols as in
%Figure~\ref{f1}. 
All the depleted BSS are located in a relatively narrow strip in the
low-luminosity part of the BSS distribution (a K-S test gives a $1\%$
probability that the luminosities of normal and depleted BSS are extracted
from the same distribution).  {\it Does this evidence indicate a significant
mass difference between CO-depleted and normal BSS?} In order to investigate
this possibility we have derived a ``photometric'' estimate of the BSS masses
by comparing their position in the CMD with a set of isochrones from
\citet[][see Figure~\ref{f2}]{cari03}. 
%The isochrone at 12 Gyr
%nicely reproduces the main cluster population \citep{cari03}.
%The isochrone at 12 Gyr (solid line) from the same dataset is also shown to
%nicely reproduce the main cluster population. 
A global metallicity $Z=0.004$ and a distance modulus $ (m-M)_V=13.36$
\citep{ferraro99} have been adopted, and the entire dataset of available
isochrones (stepped at 0.5 Gyr) has been used to derive a grid linking the
BSS colors and magnitudes to the masses.  The result is shown in the {\it
upper panel} of Figure~\ref{f3}, and reveals that the selected BSS have masses
ranging from $\sim 0.95$ to $\sim 1.35 \,M_{\odot}$, with a peak at 1.1
$M_{\odot}$. The mass distribution of the CO-depleted BSS (gray histogram) is
in reasonably good agreement with the normal-BSS histogram, and a K-S test
indicates a marginal ($\sim 2 \sigma$) difference in mass between the two
groups.

%Ours is not the first detection of abundance anomalies in
%BSS. Compared to other dwarfs, C and O have been found to be
%significantly depleted in two bright, hot BSS in M67
%\citep{mathy91}. Presumedly these are classical BMT-BSS. No such
%anomalies have been found in four cooler BSS in the same cluster
%\citep{she00}. 
%\citet{sivarani04} have recently found anomalous
%overabundances of C and N on the surface of a low-metallicity BSS in
%the halo and have interpreted these as evidence for mass-transfer
%process where the donor is an AGB star.  

High-resolution spectra also offered the opportunity to determine rotational
velocities, via the direct estimates of the Full Width at Half Maximum (FWHM)
of the measured absorption lines. Once obtained an average FWHM for each BSS,
the calibration in terms of projected rotation velocity ($V_{\rm rot} \sin i$)
has been performed by using a subsample of 4 BSS (namely 200690, 701140,
100102, 223870) observed at high resolution during the same exposures (Sabbi
et al. 2006). For these 4 BSS the procedure described in \citet{luc03} was
followed: we used their star template (observed with the same UVES set-up)
and adopted their equation (3) to derive the final values of $V_{\rm rot} \sin
i$. Under the assumption that the atmospheric parameters values do not vary
significantly within the BSS sample, we adopted this calibration to convert
the FWHM into rotational velocities and obtained 
%This process has yielded 
the largest sample of rotation velocities
ever measured for BSS in GCs. The result is shown in the {\it lower
panel} of Figure~\ref{f3}. The distribution of BSS rotation velocities
peaks at $V_{\rm rot} \sin i=7\, {\rm \,\kms}$, and shows a tail
toward larger values. The peak value is consistent with that obtained
by \citet{luc03} for a sample of TO and sub-giant branch stars in 47
Tuc ($V_{\rm rot} \sin i \sim 4\, {\rm \,\kms}$), suggesting that most
BSS are slow rotators. The high-rotation tail of the distribution
consists of 10 objects with $V_{\rm rot} \sin i > 10\,\kms$, with only
one (namely 700912) having a really large rotation velocity 
($V_{\rm rot}\sin i\sim 100\,\kms$).  
No correlation has been found between
CO-depletion and rapid rotation: only two (moderately-)rapid rotators
(602080  and 102835 with $V_{\rm rot}\sin i\sim 21\,$ and $13 \kms$,
respectively) show CO-depletion. However there is an interesting
correlation between rotation and the location of the star on the
CMD. In Figure~2 the stars with $V_{\rm rot}\sin i > 10\,\kms$ are
indicated with crosses. {\it All of the stars in the high rotation
tail lie in lower part of the BSS distribution in the CMD. }
Considering projection effects all of the BSS in this band could be
rotating at these moderately high velocities.

All the BSS in our sample display significantly lower
rotation velocity than the four BSS recently measured by
\citet{demarco05}, but unfortunately we do not have any object in
common with them, hence no direct comparison can be done. Moreover,
all their objects are located in the innermost region of the cluster
(within $20\arcsec$), while only an intermediate-rotator ($V_{\rm rot}
\sin i \sim 13\, \kms$) BSS in our sample (102835) is found
within this distance from the cluster center.  Therefore, a strong
segregation of rapid rotators toward the cluster center is needed to
reconcile the measurements obtained for the two samples.  Curiously,
%beside the low level of significance ($\sim 1\sigma$) of the
%difference between the radial distributions of rapid and ``normal''
%rotator BSS, 
the most rapid rotators in our sample (with $V_{\rm rot}
\sin i > 20\, \kms$) are located in the outer
region of the cluster, at $r>4\farcm 7$ from the center, and the
fastest rotator (BSS 700912) is at almost $9\arcmin$.  
 
\section{Discussion} 
\citet{map04} have presented a scenario explaining the radial
distribution of BSS in 47~Tuc as a mixture of C-BSS and BMT-BSS. How
do our results relate to their hypothesis?  Abundance anomalies should
provide a clear signature differentiating C-BSS from
BMT-BSS. Hydrodynamic simulations \citep{lombardi95} have shown that
very little mixing is expected to occur between the inner cores and
the outer envelopes of the colliding stars and the resulting 
 BSS should show no abundance anomalies. On the
other hand, in BMT the mass being transferred eventually comes from
deep in the donor star where partial CNO processing has occurred, thus CNO
anomalies should be expected \citep{sarna96}.

Since close binary stars often show variability, we sought to bring
additional information into our analysis by making a cross-correlation
between catalogs of variable stars and our sample.  We used the
comprehensive catalogue of variable stars of
\citet{albrow01}
%\footnote{The absolute positions listed by
%\citet{albrow01} have a systematic offset of $\sim 4\arcsec$ in RA
%with respect to our coordinates.} 
in the central regions of 47~Tuc and
that of \citet{weldrake04} in the exterior.  Three BSS presented here
 are photometric binaries: 800267 and
700912 are V6 and V14 in the \citet{weldrake04} compilation (they also
correspond to OGLEGC 227 and 250 in the Kaluzny et al. 1998
list), while 102835 is PC1-V10 of the \citet{albrow01} sample (it is
also the weak X-ray source 266 of Heinke et al. 2005). All of them
display W Ursae Majoris (W UMa)-type light curves. In the field W~UMa
objects are binary systems losing orbital momentum because of magnetic
braking. These shrinking binary systems, initially detached, evolve to
the semi-detached and contact stages (when mass-transfer starts) and
finally merge into a single star \citep{vilhu82}. In dense cluster
environments stellar interactions can drive 
binaries toward merger: these systems could reasonably be expected to display W~UMa
characteristics, although the evolutionary time scales could be very
different. The connection of W~UMa stars to BSS suggests that the BMT
channel of BSS be split into two subclasses: {\it (i)} the classical case in
which the mass transfer is driven by the evolution of the initial
primary off of the MS; {\it (ii)} the W~UMa case where the mass transfer is driven
by magnetic braking or stellar interactions.

The W~UMa stars in our sample have quite short periods (0.38, 0.35 and
0.43 days, respectively) and have been classified as contact or
semi-detached binaries.  The secondary variations observed in the
light curve of 800267 (V6) by \citet{weldrake04} and the inferred mass
ratio of 102835 (PC1-V10) by \citet{albrow01} have been interpreted as
evidence that mass transfer is active in these systems.  Two of the
three W~UMa stars are CO-depleted BSS. Unfortunately no abundance
 was derived for the rapid rotator 700912, but we can
reasonably expect that even this object might show evidence of
CO-depletion. In these W~UMa stars we appear to have caught 
the BSS mass-transfer in progress
further supporting the evidence that {\it our observations have detected the 
chemical signature of the BMT-BSS formation channel}.

In the early stage of mass transfer in W~UMa ({\it Stage-1}), the transfered
mass could come from the unprocessed material and the resulting star would
have normal C-O abundances. As the transfer continues reaching into the
region of CNO processing, first C and then both C and O would appear depleted
({\it Stage-2}). Thus it is possible to find depleted C, normal O BSS/W~UMa
stars, like 800267. After the merger the star would appear as CO depleted
non-variable BSS ({\it Stage-3}). In our sample we have found 2 or 3 stars in
{\it Stage-2}, and 4 in {\it Stage-3}\footnote{A wide range of anomalies
could be possible depending of the mass and evolutionary state of the donor
star: e.g. \citet{sivarani04} have found anomalous overabundances of C and N
on the surface of a low-metallicity BSS in the field and have interpreted
these as evidence for mass-transfer process where the donor is an AGB star.}.
Classical BMT would also result in BSS with CO depletion, perhaps with a low
mass He white dwarf companion. Since the donor star is evolving off the MS,
the transfered mass might be more heavily processed. The resulting BSS would
be very similar to a {\it Stage-3} W~UMa BSS.

The number of BSS with CO depletion and the W~UMa systems show that
the BMT channel is active even in a high-density cluster like
47~Tuc. At least 10--20\% of the BSS are being produced by the BMT
channel. This finding is in good agreement with the results of
dynamical simulations \citep{map04} which have shown that a
significant contribution of BSS ($25\%$) generated by the natural
evolution of primordial binaries is needed in order to reproduce the
bimodal radial distribution of the BSS in this cluster
\citep{ferraro04}.

%While the current result might seem to be a confirmation of the \cite{map04}
%hypothesis, things are not so simple.\footnote{Why, oh why, don't we
%ever seem to find a simple result?} The problem arises from our
%inability to adequately observe the cluster center and the resulting
%bias in the spectrscopic sample. Roughly 10\% of our sample is in the
%inner 100\arcsec\ where we we expect mostly C-BSS. 65\% of the sample
%is from the broad minimum in $N_{rm BSS}/N_{HB}$ between 100\arcsec\
%and 500\arcsec. Most of these should be BSS resulting from binaries
%migrating inward. Outside 500\arcsec\ where $N_{rm BSS}/N_{HB}$ rises
%most BSS should be BMT-BSS. Given the sample selection bias, we expect
%90\% BMT-BSS, so the vast majority of our sample should show C-O
%depletion. 
However, 
%in our sample the BMT-BSS percentage could be significantly 
%larger than this, since only $\sim 10\%$ of  the observed BSS are in the
%inner 100\arcsec\  where we expect mostly C-BSS.
%65\% of the sample
%is from the broad minimum in $F_{BSS}$  between 100\arcsec\
%and 500\arcsec observed by  Ferraro et al (2004).
%Most of these should be BSS resulting from binaries
%migrating inward. Outside 500\arcsec\ where $F_{BSS}$ rises
%most, BSS should be BMT-BSS. 
the vast majority (90\%) of BSS in our spectroscopic sample is located in the
external region of the cluster.  Hence, accordingly with Mapelli et al.
(2004, 2006), they should mainly be BMT-BSS (since C-BSS are expected to be
strongly segregated in the cluster centre). Thus,
%90\% BSS in our sample to be BMT-BSS, so 
most of them should show CO-depletion, at odd with what observed.  

However, from their relative location in the CMD, CO-depleted BSS appear to
be less evolved than non-depleted ones, possibly indicating that their
following evolution converts C-O abundances back to normal. Certainly, once
C and O have been processed into N producing CO-depletion, further nuclear
processing would not restore normal C-O abundances during the BSS phase.
Instead, mixing processes could play a role in this game.

Indeed, the distribution of rotational velocities provides a clue.  Most BSS
in our sample are slow rotators, with velocities compatible with those
measured in unperturbed TO stars \citep{luc03}\footnote{In particular, W~UMa
systems are expected to be rapid rotators. Yet, among the three BSS
identified as W~UMa systems we have found a rapid rotator (700912) and two
intermediate-slow rotators (namely 102835 and 800267). Perhaps this arises
because we are seeing the systems at different inclination angles; perhaps
not.}.
%Both formation
%channels (BMT and collision)should leave a signature of rotation on
%the newly formed BSS (according to Leonard \& Livio 1995 {\bf Does this ref
%  say that?}). 
From their location in the CMD, all of our more rapidly rotating BSS are
presumedly the most recently born. This is also the region of C-O depletion
and the W~UMa behavior. The cooler, older BSS rotate more slowly and have
``normal'' C-O. One would expect rotation to slow, and, as that happens,
mixing might be induced. Ordinarily rotational mixing increases CNO
anomalies. However in BMT-BSS C-O depleted material overlies material with
normal C-O. The result of mixing would be to push C-O back toward
"normalcy". C and possibly O would still be low, but less so than a BMT-BSS
at birth. Indeed, we do find that the bulk of our sample has C roughly one
half of that of the TO stars. {\it Is this the true origin of the offset in
the C-abundance detected in Figure 1?}
%This could be due to systematic error; it could also be a
%real effect. It is crucial that we understand which.
 
\acknowledgements  
This research was supported by contract ASI-INAF I/023/05/0 and 
the Ministero dell'Istruzione, dell'Universit\`a e della Ricerca.
We warmly thank Alessio Mucciarelli for a number of checks he performed on
the data.

\clearpage

\clearpage

\clearpage

\end{document}